\documentstyle[twoside,12pt]{article}
\newcommand{\bp} {{\bf p}}

\newcommand{\bsigma} {{\mbox{\boldmath $\sigma$}}}

\newcommand{\bnabla} {{\mbox{\boldmath $\nabla$}}}
\newcommand{\btab}{\begin{tabbing}}
\newcommand{\etab}{\end{tabbing}}

\newcommand{\beqn}{\begin{equation}}
\newcommand{\eeqn}{\end{equation}}
\newcommand{\barr}[1]{\begin{array}{#1}}
\newcommand{\earr}{\end{array}}
\newcommand{\beqna}{\begin{eqnarray}}
\newcommand{\eeqna}{\end{eqnarray}}
\newcommand{\btablec}{\begin{table} \begin{center}}
\newcommand{\etablec}{\end{center} \end{table}}
\newcommand{\lapprox}{\stackrel{<}{\scriptstyle \sim}}

\newcommand{\gapproxeq}{\lower.7ex\hbox{$\;\stackrel{\textstyle>}{\sim}\;$}}
\newcommand{\lapproxeq}{\lower.7ex\hbox{$\;\stackrel{\textstyle<}{\sim}\;$}} 
\newtheorem{theorem}{Theorem}
\newcommand{\bth}{\begin{theorem}}
\newcommand{\eth}{\end{theorem}}

\newcommand{\br}{{\bf r}}

\newcommand{\rf}{{\varsigma}}
\newcommand{\rff}{{\varrho}}
\newcommand{\ra}{{\varsigma_1}}
\newcommand{\rb}{{\varsigma_2}}

\newcommand{\plabel}[1]{\label{#1}}
\newcommand{\pbibitem}[1]{\bibitem{#1}}
\newcommand{\sr}{{\cal R}}
\newcommand{\gr}{\gamma_{\rho}}
\newcommand{\pre}{\frac{\pi\alpha}{\gr^2}}
\newcommand{\presq}{(\frac{\pi\alpha}{\gr^2})^2}
\newcommand{\sd}{{\cal D}}
\newcommand{\sff}{{\cal F}}
\newcommand{\sm}{{\cal M}}
\newcommand{\sg}{{2\beta^2_A+\beta^2}}
\newcommand{\uf}{{f}}
\input epsf

\begin{document}
\title{\begin{flushright} \small{hep-ph/9607475}\\
\small{MC-TH-96/20} \end{flushright} 
\vspace{0.6cm}  
\LARGE \bf Two Photon Couplings of Hybrid Mesons}
\author{Philip R. Page\thanks{\small \em E-mail: prp@jlab.org.} \thanks{\small \em Present address:
Theory Group, Thomas Jefferson National Accelerator Facility, 
12000 Jefferson Avenue, Newport News, VA 23606, USA.}\\
{\small \em Department of Physics and Astronomy, University of Manchester,} \\
{\small \em Manchester M13 9PL, UK}}
\date{August 1996}
\maketitle
\abstract{A new formalism is developed for the two photon production 
of hybrid mesons via intermediate hadronic decays. 
In an adiabatic and non--relativistic context with spin 1 $Q\bar{Q}$ pair
creation we obtain the first
absolute estimates of unmixed hybrid production strengths to be small ($\; \lapprox
0.03 - 3$ eV)
in relation to experimental meson widths ($\sim 0.1 - 5$ keV). 
Within this context, $\gamma\gamma$ experiments at Babar, Cleo II, LEP2
and LHC therefore strongly discriminate
between hybrid and conventional meson wave function components, filtering out 
conventional meson components.
Decay widths of unmixed hybrids vanish.
Conventional meson two photon widths are 
roughly in agreement with experiment.}

\vspace{2cm}

\normalsize

Due to the building of ever higher energy accelerators with a consequent
increase in quasi--real photon emission \cite{sadovsky95,baur95}, the probability for resonance
production  
via two photon collisions becomes significant \cite{feindt95}. This can open up a promising
new pathway whereby new forms of matter with an explicit excitation of
the gluonic degree of freedom can be produced. These ``gluonic''
hadrons are predicted \cite{amsler96,perantonis90} by QCD and their 
discovery represents an important check of the Standard
Model. One such hadron
for which there is preliminary evidence \cite{page95light,page96rad,page96panic} is excitations in the
presence of $Q\bar{Q}$ systems (called ``hybrid mesons'').
Obtaining a first estimate of the absolute two photon decay widths and 
production cross--sections
of hybrid mesons forms the motivation for what follows.

In order to make an {\it absolute} prediction, 
all calculational parameters need to be fixed by known experimental 
observables or theoretical models, which is difficult given that no unambiguous hybrid
meson candidate has been found so far. However, we know the probability of vector mesons coupling to
photons from the
$e^+e^-$--widths of vector mesons. If we incorporate the  hadronic decays of hybrid
mesons into two vector mesons,
which have been calculated in the flux--tube model \cite{page95light},
we can predict two photon production strengths of hybrid mesons \cite{babcock76}. This picture of an intermediate
hadronic ``kernel'' in two photon decays of hybrids must happen at some
level in nature. Moreover, there are theoretical reasons why such a
treatment is needed. In heavy quark lattice gauge theory \cite{perantonis90} and model realisations
of it (e.g. the flux--tube model \cite{paton85}), the interquark flux--tube
is excited with non--zero angular momentum around
the $Q \bar{Q}$ axis. In such a picture it is not clear how direct coupling of two photons  
to hybrids can be achieved, due to the inability of the photons to carry off the non--zero
angular momentum.

There is a straightforward way to couple two photons to vector meson intermediate
states (as we shall see in the next section). Based on this, it can be
shown that hybrid meson two photon couplings are small. This result is
independent of detailed dynamics in the hadronic kernel. 
Henceforth we show how $\gamma\gamma$ collisions emerge as an
avid discriminator between gluonic and non--gluonic $Q\bar{Q}$
wave function components, acting as a promising process  for the isolation of new hybrid forms
of matter when used in conjunction with
hadronic production mechanisms.

The outline of the paper is as follows. In section \ref{sec1} we
introduce the formalism and couple photons
to intermediate vector mesons. From this discussion the emerging phenomenology of
section \ref{ph} can be deduced. Section \ref{sec2} demonstates that the formalism
respects general principles, including
Yang's theorem and Bose symmetry. 
In section \ref{sec3} the coupling of photons to vector mesons and the
properties of intermediate vector mesons are
quantitatively formulated. The results of including a detailed
flux-tube model hadronic kernel for conventional and hybrid meson two photon couplings
are displayed in section \ref{sec4}.

\begin{figure}
\begin{center}
\plabel{figq}
\caption{Diagrams constributing to the two photon decays of hybrid and
conventional mesons. Right arrows indicate quarks, and left arrows antiquarks.}
\leavevmode
\hbox{\epsfxsize=4 in}
\epsfbox{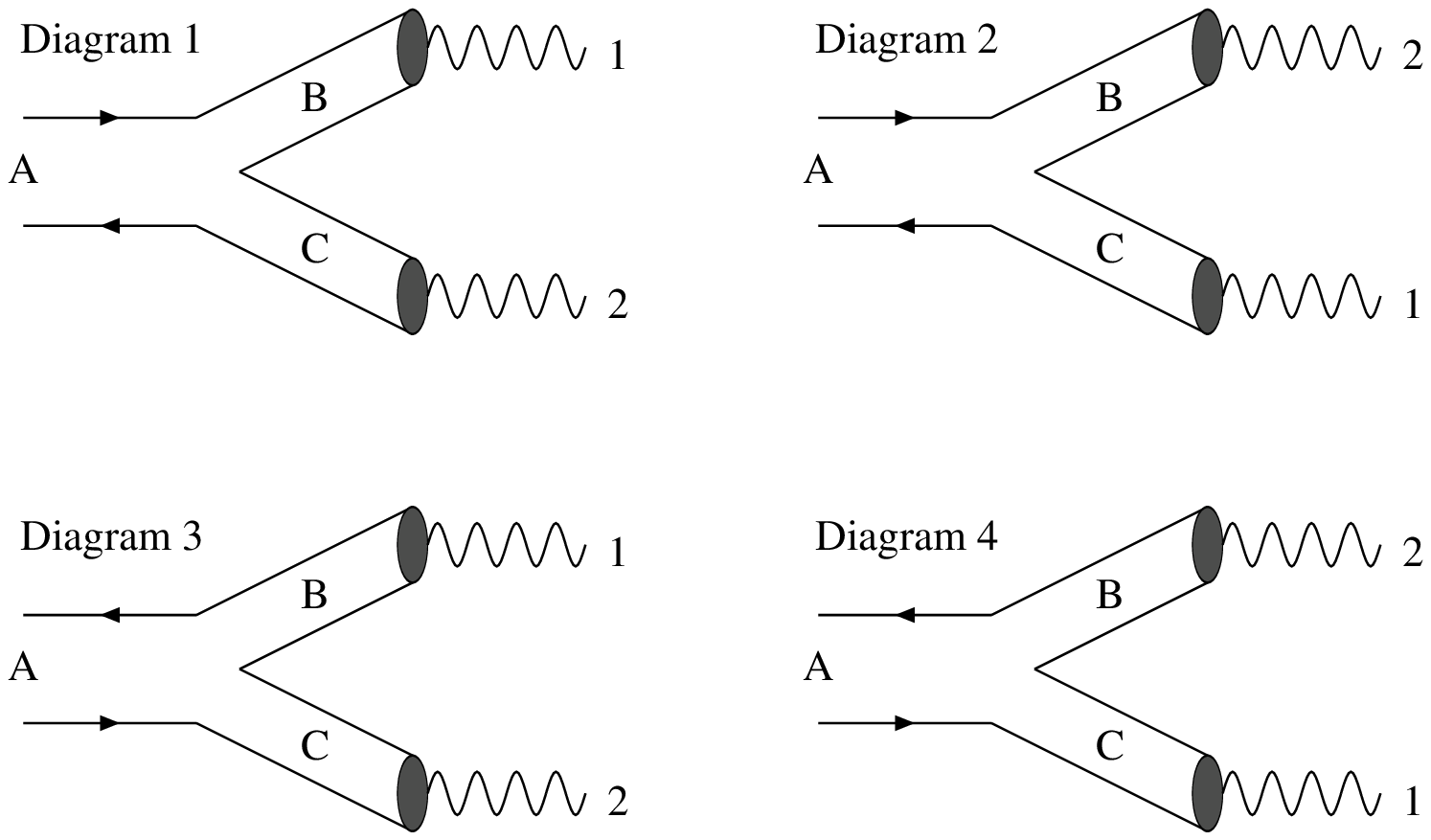}
\end{center}
\end{figure}

\section{Overall Features of the Formalism \plabel{sec1}}

We work in the rest frame of the conventional or hybrid meson A, which
couples to vector mesons B and C. These 
in turn each couples to photon 1 or 2 (see Fig. 1). 
A vector meson has a non--relativistic 
spin  projection $M_{S}^V$ equal to the angular momentum projection $M_J^i$ 
of the photon $i$. The photons are at first assumed to be off--shell
$q_{i}^2 \neq 0$, and  hence can be distinguished based on the square
of their four--momenta.
There are {\it two} ways of coupling the photons: with
vector mesons B and C coupled to photons 1 and 2 respectively (called ``Diagram 1''),
and with B and C coupled to 2 and 1 respectively (``Diagram 2''). The hadronic 
kernel is unchanged in these two diagrams, except to the extent that B
and C couple to different photons.  This means that to obtain Diagram 2 from Diagram 1 
we have to replace the momentum of vector meson B, denoted by $\bf{p_{B}}$, to $ - \bf{p_{B}}$;  and
that we have to interchange $M_{S}^{B}$ and $M_{S}^{C}$. Replacing
$\bf{p_{B}}$ by $ - \bf{p_{B}}$ introduces a sign $(-1)^{L}$ for a
decay amplitude in partial wave $L$.

Since the $Q\bar{Q}$ pair is assumed to be created with spin 1, the hadronic kernel
will contain spin dependence $\bsigma =(\sigma_x,\sigma_y,\sigma_z)$
in terms of Pauli matrices. We then take the overlap of the spin
wavefunctions of A, B and C,  and obtain
in the language of refs. \cite{page95light,page95thes} a spin dependence 
$Tr(A^{T}B\bsigma^{T}C)^{SM_S}$ in the amplitude. Here the spins of states A, B, or C are denoted
by the appropriate matrix A, B, or C. 
$Tr(A^TB\bsigma^TC)^{SM_S}$ is found to change by $(-1)^{S_A+S_B+S_C+1}$
under exchange of $M_S^B$ and $M_S^C$ (see the Appendix of
ref. \cite{sel2}). Here $S_A$ is the spin of state A and $S_B = 1 = S_C $.

Since Diagram 2 and Diagram 1 are related by a sign $(-1)^{L}$, 
and the rule for exchange of $M_{S}^{B}$ and $M_{S}^{C}$, it follows that
Diagram 2 = $(-1)^{S_A+S_B+S_C+1+L}$ 
Diagram 1. In the limit of adiabatically
moving quarks, the $Q\bar{Q}$  spin for mesons are just those of the 
non--relativistic quark model. For hybrids the $Q\bar{Q}$  spin is 1 for
$J^{PC} = (0,1,2)^{-+}$ and 0 for $1^{++}$ \cite{perantonis90,paton85}.
By explicitly considering each two photon process discussed
in this paper, we obtain that
$(-1)^{S_A+S_B+S_C+1+L}$ is 1 for mesons (constructive interference) and $-1$ 
 for hybrids (destructive interference).  
Hence for hybrids  Diagrams 1 and 2 cancel. This yields
a vanishing hybrid meson two photon coupling.

There is, however, something that has been left out of the above 
discussion. The meson
propagator contains a term that links the $q_{i}^2$ of the photon to
the vector meson mass $m_V$ (see Eq. \ref{vertex}). Each diagram is
proportional to the product of the propagators of vector meson B and C. This means
that if $q_{1}^2 \neq q_{2}^2$, the symmetry under exchange of
photon labels is broken if $m_B \neq m_C$. Hence two photon couplings
of hybrid mesons 
are not exactly zero, but small. In the case of decays the photons are
real, i.e. $q_1^2 = 0 = q_2^2$, hybrid meson two photon decay widths
are exactly zero.

In Diagrams 1 and 2, we have the quark of state A contained in 
B, and the antiquark in C. When B and C are distinguishable (which they
are when they have different masses or flavours), there is a second
possibility: the antiquark of state A can be contained in B, and the quark in
 C. These possibilities are denoted by Diagrams 3 and 4.

The same reasoning applied previously to the relationship between
Diagrams 1 and 2 can also be applied to the relationship between
Diagrams 3 and 4. We conclude that hybrid meson two photon couplings are
small and decay widths vanish.
This is the central claim of this paper.

\section{Consistency Checks on the Formalism \plabel{sec2}}

\vspace{.1cm}

\noindent {\bf Yang's Theorem}

\noindent Yang's theorem states that total angular momentum $J_A=1$ states do not couple to
two real photons. We explicitly check that this is satisfied. 

Real photons can only be transversly polarized,
so that $M_S^V = \pm 1$. For a $1^{++}$ meson this means that only
$M_S^A=0$ contributes. In addition, it can be shown that 
$Tr(A^{T}B\bsigma^{T}C)^{SM_S} = 0$ for $M_S^A=0,\; M_S^B=\pm 1,\; M_S^C=\mp
1$
if A has spin 1. Because a $1^{++}$ meson has spin 1
\cite{perantonis90,paton85}, it follows that it does not couple to real
photons, as required.  
A $1^{++}$ or $1^{-+}$ hybrid does not couple to real photons because
all hybrid amplitudes vanish in this case.

By postulating that the coupling of a longitudinal photon to a vector meson is proportional
to $f_{i}$ (Eq. \ref{vertex}), where $\uf_i \; (q_i^2=0) =0$, we can
insure that real
photons cannot be longitudinally polarized, and that Yang's theorem 
is lifted when $q_{i}^2\neq 0$.

\vspace{.35cm}

\noindent {\bf Bose Symmetry}

\noindent When $M_J^1 = M_J^2$ and $q_1^2 = q_2^2$ the photons are
identical. By
Bose symmetry we expect their coupling to vanish if the process happens in 
odd wave, e.g. for $(0,1,2)^{-+} \rightarrow \gamma\gamma$. For
$(0,2)^{-+}$ {\it mesons}
Bose symmetry is explicitly protected by the fact that exchange $M_J^1 
\leftrightarrow M_J^2$
introduces a sign $(-1)^{S_A+S_B+S_C+1}$ which is odd (see previous
section). But we know that $M_J^1 = M_J^2$, so that exchange $M_J^1 
\leftrightarrow M_J^2$ trivially introduces an even sign. Thus the
coupling vanishes. For {\it hybrid} $(0,1,2)^{-+}$ we already know that the
amplitudes vanish when $q_1^2 = q_2^2$.

\vspace{.35cm}

\noindent {\bf Electric Charge} 

\noindent In a quark level picture of two photon decay widths \cite{ackleh92,li92}
$u\bar{u}\pm d\bar{d}\rightarrow\gamma\gamma$ is 
proportional to $((\frac{2}{3})^2 \pm (-\frac{1}{3})^2)^2$ due to the
$u,d$ electric charges, yielding a width ratio $9:25$ for isospin
$I=1\; :\; I=0$ states.
Due to the fact that flavour is treated similarly in this formalism,
we expect the same result. From Eqs.  \ref{rmes} and \ref{param} (see below)
$\; I=1:I=0 \; = \; 8 \sr_{\omega}^2 : 2 (1+\sr_{\omega}^2)^2  \; \approx \; 
9:25$, consistent with expectations.

\vspace{.35cm}

The preceding checks\footnote{The general result
\protect\cite{berger87} that $2^{-+}\rightarrow\gamma\gamma$ has no
helicity 2 components can also be verified.} are satisfied, as expected,   
independent of detailed dynamics. Having acquired confidence in the formalism,
we proceed in the following sections to display detailed model calculations.

\section{Intermediate Vector Mesons\plabel{sec3} }

The coupling strength of photon $i$ to
vector meson V is defined by the dimensionless product of the vector meson dominance vertex
and the vector meson propagator 

\beqn
\plabel{vertex}
V_i  = \frac{-em_V^2\sr_V}{2\gr}\frac{1}{q_i^2-m_V^2}
\;\left\{ \barr{cl} 1 & \mbox{transverse } i \\ 
\uf_i & \mbox{longitudinal } i  \earr \right.
\hspace{1.3cm} \sr_{\rho} = 1 
\eeqn
where $\uf_i$ is a continuous function of $q^2_i$ such that $\uf_i \; (q_i^2=0) =0$.
The constant $\gr$ can be fitted from the $e^+e^-$ width of the $\rho$ meson.  
$\sr_V$ parameterizes the ratio of the photon coupling amplitude for vector meson
$V$ relative to that of the $\rho$ meson, and can be fitted from 
the $e^+e^-$ width of $V$.

We now investigate the relationship between Diagrams 1 and 3. They are related by the quark in A being contained in either B or C
respectively. In the studies of hadronic decays \cite{page95light,kokoski87,kokoski85}
these diagrams have been shown \cite{page95thes} to be related by the sign
$(-1)^{I_A+I_B+I_C+S_A+S_B+S_C+1+L+\Lambda}$, which always equals 
unity for allowed couplings. This implies that Diagrams 1 and 3
(and similarly Diagrams 2 and 4) are numerically identical.
The angular momentum $\Lambda$
of the flux around A's moving $Q\bar{Q}$ axis vanishes for mesons and
equals $\pm 1$ for hybrids.
The sign differs from that in section \ref{sec1}
because photons do not ``know'' about flavour and flux--tube degrees of freedom.

When B and C are not distinguishable (in mass and flavour), both 
Diagrams 1 and 3 are included in conventional studies
of hadronic decay \cite{page95light,kokoski87,kokoski85} since 
B and C can be distinguished based on their momenta. However, when B and
C are intermediate states, they cannot be distinguished in this way.
Hence only Diagrams 1 and 2 are included in this calculation for 
intermediate states $\rho^0\rho^0,\; 
\omega\omega,\; \phi\phi,\; \psi\psi$, and not Diagrams 3 and 4. 

When we calculate the square of the amplitude for the diagrams in
Fig. 1, we obtain a common term which is independent of
the $J^{PC}$ of state A, denoted by

\beqn
 \plabel{df}   
\sd_{\pm}  \equiv  (Tr(AB^TC^T)^{F} \; \frac{1}{2}(\mbox{Number of Diagrams})\;  (\sum_{B,C}
B_1 C_2 \pm B_2 C_1))^2
\eeqn
where we group 
together the flavour overlap $Tr(AB^TC^T)^{F}$, the various diagrams
and the photon couplings. The summation in  Eq. \ref{df} refers to the sum over
 intermediate hadronic states B and C. The term inside the summation
 refers to the
 product of the $V_i$ for Diagram 1 added or subtracted to the product of the $V_i$
 for Diagram 2. For conventional mesons we have addition and for
 hybrid mesons subtraction as discussed in section  \ref{sec1}. If Diagrams 3
 and 4 contributes, their contribution is the same as that of Diagrams 1
 and 2, and is incorporated via the ``Number of Diagrams'' term in
 Eq. \ref{df}. 

From Eq. \ref{vertex}
\beqna \plabel{pl} 
\sum_{B,C} B_1 C_2 + B_2 C_1 & = & 2 \pre \sum_{B,C} \sr_B \sr_C \\  
\sum_{B,C} B_1 C_2 - B_2 C_1 & = & \pre
\sum_{B,C}\sr_B\sr_C\sff_{BC}\nonumber \\  
\mbox{where } \sff_{BC} & \equiv &  ((1-\frac{q_1^2}{m_B^2})(1-\frac{q_2^2}{m_C^2}))^{-1} - ((1-\frac{q_2^2}{m_B^2})(1-\frac{q_1^2}{m_C^2}))^{-1}\plabel{fed} \eeqna 
and we simplified Eq. \ref{pl} for real photons. 
Since $\sff_{BC}=0$ for real photons, hybrid meson widths vanish. 
Equally,
when $q_1^2 = q_2^2$ the production amplitudes vanish. Hence hybrids are 
expected to be produced only if $q_1^2 \neq q_2^2$, and even then
with at least three orders of magnitude suppression (see Fig. 2 and section \ref{sec4}).

For meson two
photon decays, the low--lying (G--parity allowed) hadronic 
intermediate states are included as

\beqna \plabel{rmes}
I=1: & \rho\omega & \sd_+ = (\frac{1}{\sqrt{2}} \;  2 \; 2 \pre \sr_{\omega})^2 = 
8 \presq \sr_{\omega}^2 \nonumber \\ 
I=0: & \rho\rho, \; \omega\omega & \sd_+=(\frac{1}{\sqrt{2}} \;  1  \; 2\pre
(1+\sr_{\omega}^2))^2 = 2 \presq (1+\sr_{\omega}^2)^2  \nonumber \\
s\bar{s}: & \phi\phi & \sd_+=(1\; 1  \; 2\pre \sr_{\phi}^2)^2 = 4 \presq \sr_{\phi}^4 \nonumber \\ 
c\bar{c}: & \psi\psi &  \sd_+=(1\; 1  \; 2\pre \sr_{\psi}^2)^2 = 4 \presq \sr_{\psi}^4
\eeqna
where each of the components of $\sd_+$ in Eq. \ref{df} is explicitly
indicated.
For two photon decays of hybrids the dominant intermediate states are included as  

\beqna \plabel{hybr}
I=1:& \rho_R\omega,\; \omega_R\rho & \sd_{-} = (\frac{1}{\sqrt{2}} \;  2  \; \pre
(\sr_{\rho_R}\sr_{\omega}\sff_{\rho_R\omega}+\sr_{\omega_R}\sff_{\omega_R\rho}))^2 
 \nonumber \\
I=0:& \rho_R\rho,\;  \omega_R\omega & \sd_{-}=(\frac{1}{\sqrt{2}} \;  2  \; \pre 
 (\sr_{\rho_R}\sff_{\rho_R\rho}+\sr_{\omega_R}\sr_{\omega}\sff_{\omega_R\omega}))^2 \nonumber \\
s\bar{s}:& \phi_R\phi &\sd_{-}=(1\; 2  \; \pre \sr_{\phi_R}\sr_{\phi}\sff_{\phi_R\phi})^2\nonumber \\ 
c\bar{c}:& \psi_R\psi &\sd_{-}=(1\; 2  \; \pre \sr_{\psi_R}\sr_{\psi}\sff_{\psi_R\psi})^2
\eeqna
where the minus sign in $\sff_{BC}$ explicitly incorporates the
destructive interference derived in 
section \ref{sec1}.
We choose the intermediate vector mesons in Eq. \ref{hybr}, involving
radially excited states $\rho_R,\; \omega_R,\; \phi_R$ and $\psi_R$, for
the following reasons. 
When $m_B \approx m_C$, as is the case for hybrid decays to two low--lying 
S--wave vector mesons, $\sff_{BC} \sim m_B - m_C$ is small,
suppressing the decay amplitude for  
these intermediate states by $\lapprox \frac{m_{\rho}-m_{\omega}}{m_{\rho_R}-m_{\omega}} = 2 \%$
relative to the modes listed in Eq. \ref{hybr}.
In addition, for hybrid decays into low--lying S--wave vector mesons,
the hadronic kernel is proportional to 
$(\frac{\beta_B^2-\beta_C^2}{\beta_{B}^2+\beta_{C}^2})^2$ in the flux--tube model \cite{page95light}
when S.H.O. wave functions with inverse radii $\beta_B$ and $\beta_C$ are used. This is zero for $\rho^0\rho^0,\; \omega\omega,\; 
\phi\phi,\; \psi\psi$ and $\lapprox 10^{-4}$ \cite{page95thes} for $\rho\omega$.
Dominant contributions to two photon widths of hybrids are thus expected
to come from intermediate states consisting of 
one radially excited (2S), D--wave or hybrid meson with one low--lying
S--wave meson.
Hybrid meson coupling to a photon is generally 
\cite{ono84} believed to be suppressed, and at least in the non--relativistic
limit, D--wave mesons are also coupled weakly to photons\footnote{
With relativistic corrections $\sr_{\rho_{D}} = 0.10$, $\sr_{\omega_{D}} = 
0.030$, $\sr_{\phi_{D}} = 0.061$  \cite{isgur85}.}. Hence the choice of terms 
in Eq. \ref{hybr} represents dominant intermediate states.

\subsection{Parameters}

Using the fact that the $e^+e^-$ width of a vector meson is
$\frac{\pi\alpha^2}{3} \frac{m_V}{\gamma_{\rho}^2} \sr_{V}^2$, 
we can obtain $\frac{\gamma_{\rho}^2}{4\pi} = 0.507$ and
\beqn \plabel{param} \sr_{\omega} = 0.30 \; \; \; \sr_{\phi} = 0.39 \; \; \; 
\sr_{\psi} = 0.44 \;\; \;  \sr_{\psi_R} = 0.26 \eeqn
using  experimental $e^+e^-$ widths and masses \cite{pdg94}.
For the higher mass vector mesons, similar considerations using the $e^+e^-$ widths of ref. \cite{donnachie94} show that the values of $\sr_{V}$ derived 
are in substantial disagreement with theoretical expectations \cite{isgur85}
(see Eq. \ref{rrad}) which may arise due to substantial mixing between 2S, 1D and hybrid in the physical states \cite{page96rad,donnachie94}.
Relativized quark models predict\footnote{Using $\sr_V =
f_V/f_{\rho}$, where $f_V$ is defined in ref. \protect\cite{isgur85}.} that
\beqn \plabel{rrad} \sr_{\rho_R} = 0.19 \;\; \;  \sr_{\omega_R} = 0.061 \;\; \;  \sr_{\phi_R}=0.14 \eeqn
which we adopt.

\section{Flux--tube Model Hadronic Kernel and Results \plabel{sec4} }

We adopt a
hadronic decay model which predicts hybrid meson decays \cite{page95psi} by fixing
parameters from known conventional meson decays, called the 
non--relativistic flux--tube
model of Isgur and Paton \cite{paton85}. In addition to providing absolute
estimates for the hadronic kernel, this model reduces for
S.H.O. wave functions \cite{page95psi} to the phenomenologically successful \cite{geiger94}
$^3 P_0$ decay model. 

\subsection{Meson Coupling}

\begin{table}[t]
\begin{center}
\caption{Two photon theoretically predicted and experimentally
observed \protect\cite{pdg94} widths (in keV) of the lowest radially 
excited mesons of various
$J^{PC}$ and flavour. When masses are not listed in ref. 
\protect\cite{pdg94}, their
assumed values are listed (in MeV) in square brackets. 
Due to details of phase space conventions hadronic decay models can
also allow $\gamma_0 = 0.53$ \protect\cite{geiger94}, and hence the
theoretical predictions listed can be $(0.53/0.39)^2 \approx 2$ times bigger.
Light pseudoscalars are well described \protect\cite{berger87} by
chiral dynamics and are not quoted. 
The broad nature of the $f_0 (1370)$ has not been taken account of. 
We took $\beta_A =$ 
0.33, 0.40, 0.50, 0.31, 0.39, 0.47, 0.45, 0.57 GeV for 
$\{ a_2,a_0,f_2,f_0(1370)\},\; 
\{ f_2^{'},f_0(1710),\eta_2(1875)\},\;  
\{\chi_{c2},\chi_{c0}\},\;  \pi_2,\;  \pi, \;
\{\eta_{u\bar{u}},\eta_{s\bar{s}}\},\;  \eta_{c2},\;  \eta_{c}$. Also
$\beta= $ 0.39, 0.47, 0.57 GeV for $u\bar{u},\;  s\bar{s},\;  c\bar{c}$. 
Theory error estimates are obtained by varying $\beta_A,\beta$ 
in the same direction within $0.05$ GeV of the mean values above. 
$\star$ Ref. \protect\cite{crys94}.
$\P$ Ref. \protect\cite{argus96}. 
$\dagger$ Ref. \protect\cite{bugg96}. 
$\ddagger$ Ref. \protect\cite{fulton95}. 
$\amalg$ Ref. \protect\cite{isgur85}.} 
\label{table1}
\begin{tabular}{|c|c||l|l|l|l|}
\hline 
 \multicolumn{2}{|c||}{} & \hspace{.4cm} $0^{-+}$& \hspace{.4cm} $0^{++}$& \hspace{.4cm} $2^{++}$& \hspace{.4cm} $2^{-+}$\\
\hline \hline 
          &     & \boldmath{$\pi$}                     & \boldmath{$a_0 [1450]$} $^\star$ &  \boldmath{$ a_2           $} & \boldmath{$\pi_2(1670) $} \\
$I=1$     & Th  & $ - \;                    $ & $ 0.7 \pm 0.5             $ & $ 0.9 \pm 0.1          $ & $ 0.1 {\lower.7ex\hbox{$\;\stackrel{\textstyle +0.3}{-0.1}\;$}}$ \\
          & Ex  & $ 7.29\pm 0.19 \; \mbox{eV}          $ & $ -                $ & $ 1.04\pm 0.09  $ & $ 1.13\pm 0.24 $ \\
          & Ex  & $       $ & $ $ & $0.96\pm 0.13$ $\P   $ & $ < 0.19 $ $\P $ \\
\hline 
          &     & \boldmath{$      $} & \boldmath{$f_0 (1370)       $} &  \boldmath{$ f_2           $} & \boldmath{$\eta_2 [1645]$} $^\dagger $ \\   
$I=0$     & Th  & $                                 $ & $ -                $ & $ 2.8 \pm 0.2             $ & $ 0.4{\lower.7ex\hbox{$\;\stackrel{\textstyle +0.7}{-0.3}\;$}} $ \\
          & Ex  & $                      $ & $ 5.4\pm 2.3       $ & $ 2.8\pm 0.4    $ & $ - $ \\
\hline 
          &     & \boldmath{$     $} & \boldmath{$ $} &  \boldmath{$ f_2^{'}       $} &  \\
$s\bar{s}$& Th  & $                                 $ & $             $ & $ 0.17\pm 0.01           $ & $  $ \\
          & Ex  & $                        $ & $                 $ & $ 0.105\pm 0.017 $ & $  $ \\
\hline 
          &     & \boldmath{$ \eta_c                    $} & \boldmath{$ \chi_{c0}        $} & \boldmath{$\chi_{c2}     $} & \boldmath{$\eta_{c2} [3840]$} $^\amalg $ \\
$c\bar{c}$& Th  & $ 7.2\pm0.2                                 $ & $ 4.1\pm0.4                $ & $ 2.2\pm0.1             $ & $ 2.0 \pm 0.3 $ \\
          & Ex  & $ 7.5\pm 1.5                          $ & $ 4.0\pm 2.8       $ & $ 0.37\pm 0.17$ & $ - $ \\
          & Ex  & $ 4.3\pm 1.4$ $^{\ddagger}            $ & $ 1.7\pm 0.8$ $^{\ddagger} $ & $ 0.7\pm 0.3$  $^{\ddagger}$& $  $ \\
\hline 
\end{tabular}
\end{center}
\end{table}

\noindent The analytic expressions for a meson coupling to two vector mesons
in the flux--tube model, with 
$^3 P_0$ pair creation dynamics and non--relativistically moving quarks,  
is identical \cite{page95psi} to the $^3P_0$ model \cite{kokoski87} in
the case of S.H.O.
wave functions. This is true if we make the identification $\gamma_0 = \frac{a \tilde{c}}{9\sqrt{3}} \frac{1}{2} A^{0}_{00} 
\sqrt{\frac{fb}{\pi}} \; (1+\frac{fb}{2\beta^2})^{-1}$ for the $^3
P_0$ model pair creation
constant, where the flux--tube model constants $f=1.1, \; b=0.18 \;
GeV^2$, $A^{0}_{00}=1.0$, $\tilde{c}$ and $a$ are defined in
refs. \cite{page95light,page95psi,perantonis87}. It is also assumed
that $\beta_B
= \beta_C \equiv \beta$.
We set $\gamma_0 = 0.39$ for $\beta=0.4$ GeV, in accordance with meson decay phenomenology
\cite{kokoski87,geiger94}.
The hadronic kernel is 
completely specified, and can be calculated using ref.  \cite{page95psi}. 
The sum over total angular momentum projection of the
squares of the helicity amplitudes $\sum_{M_J^A,M_S^B,M_S^C} |\sm
|^2$ is (for various $J^{PC}$)

\beqna \plabel{mes}
0^{-+} & : & \hspace{0.3cm} \rf\; (\bp_B(\beta^2_A+\beta^2))^2 \nonumber  \\  
0^{++} & : & \hspace{0.3cm} \rf\; \frac{1}{3}(\frac{\beta_A}{\sg})^2\;  [2(\ra-\rb)^2+\uf_1^2\uf_2^2(\ra+\rb)^2] \nonumber  \\  
1^{++} & : & \hspace{0.3cm} \rf\; (\frac{\beta_A}{\sg})^2 (\uf_1^2+\uf_1^2)\rb^2\nonumber\\  
2^{++} & : & \hspace{0.3cm}  \rf\; \frac{1}{3}(\frac{\beta_A}{\sg})^2\;  [28 \ra^2 - 8\ra\rb + 4 \rb^2
+(3(\uf_1^2+\uf_2^2)+2\uf_1^2\uf_2^2)\; (2\ra-\rb)^2] \nonumber\\  
2^{-+} & : & \hspace{0.3cm}  \rf \;\frac{4}{3} (\frac{\bp_B\beta_A^2}{(\sg)^2})^2\;  [(2\ra-\rb)^2+
3(\uf_1^2+\uf_2^2)\ra^2] \nonumber\\ & &  \nonumber\\ & &   
\hspace{1cm} \rf  \equiv  \sd_{+} \; 2^8 \pi^{\frac{3}{2}} \gamma_0^2 \; \frac{\beta_A^3}{(\sg)^5} 
\exp(-\frac{\bp_B^2}{2(\sg)}) \nonumber\\ & & 
\hspace{1cm} \ra  \equiv  \beta^2\; (\sg) \hspace{1cm} \rb \equiv \bp_B^2
\; (\beta_A^2+\beta^2)
\eeqna 
where $\beta_A$ is the S.H.O. wave function inverse radius of state
A. In Eq. \ref{mes} we explicitly seperate
longitudinal and transverse contributions: Expressions with
one $\uf_i^2$ correspond to photon $i$ being longitudinal, and those
containing $\uf_1^2 \uf_2^2$ have both photons longitudinal\footnote{
The expressions for purely hadronic decays are retrieved by setting
$\uf_1^2 = \uf_2^2 = 1$ in Eqs. \protect\ref{mes} and \protect\ref{hy}}.

Using the parameters of refs. \cite{page95light,page95thes}, and with the help of
Eqs. \ref{rmes}, \ref{mes} and \ref{wid} for real photons, we obtain
the meson width predictions of
Table \ref{table1}. 
No parameter fitting has been done, and it is
encouraging that the amplitudes roughly correspond\footnote{
As in this formalism, quark models \protect\cite{ackleh92} also find small $\pi_2\rightarrow\gamma\gamma
= 0.11 - 0.27$ keV.} to experimental values, 
establishing the validity of the model.
There can, however, be substantial sensitivity if $\beta_A,\beta$ is varied, especially
near nodes in the amplitude. 

Within this paper, it is sufficient to check the overall consistency of
the approach. More detailed work would have to take into account relativistic
corrections, which are known to be substantial
\cite{ackleh92,li92} at least in quark model approaches, and will have
to fit $\gamma\gamma$ decays contingent on the corresponding hadronic $VV$ decays 
fitting experiment. The results in Table \ref{table1} complement\footnote{
Sometimes the results are different from quark models. 
Firstly, this formalism can be shown not to allow a solution where 
$0^{++}$ and $2^{++}$ meson two photon widths are $15/4$, as expected
in the na\"{\i}ve quark model, but not necessitated by experiment.
Secondly, $b\bar{b}$ two photon production is found to be negligible
compared to quark models \cite{ackleh92} due to the
large phase space suppression $\sim \exp(-\frac{q^2}{6\beta^2})$ (see Eq. \protect\ref{mes}).
} quark model \cite{ackleh92,li92}
and quark model with vector meson dominance \cite{li92} approaches. These
can be shown to be related to each other \cite{li92}, motivating vector meson
dominance at the quark level. 

\subsection{Hybrid Production}

The flux--tube model predicts the hybrid pair creation constant
$\tilde{\gamma}_0 = \kappa \sqrt{b} \gamma_0 \; (1+\frac{fb}{2\beta^2})^{-1}$
for S.H.O. meson wave functions, in terms of the meson pair creation
constant, where the model constant $\kappa=0.9$ is defined 
in refs. \cite{page95light,page95psi,perantonis87}. 
$\sum_{M_J^A,M_S^B,M_S^C} |\sm |^2$ for 
various $J^{PC}$ is (see Appendix \ref{ap2})

\beqna \plabel{hy}
0^{-+} & : & \hspace{0.3cm} 0 \nonumber  \\  
1^{++} & : &  \hspace{0.3cm} \rff\; 2 \; [4(\breve{g}_0+\breve{g}_2)^2+(\uf_1^2+\uf^2_2)(2\breve{g}_0-\breve{g}_2)^2] \nonumber\\  
1^{-+} & : & \hspace{0.3cm} \rff\; 9\breve{g}_1^2\;  [\uf_1^2+\uf^2_2+2\uf_1^2\uf^2_2] \nonumber\\  
2^{-+} & : & \hspace{0.3cm}  \rff\; 9\breve{g}_1^2\;  [8+\uf_1^2+\uf^2_2] \nonumber\\ \nonumber\\  &  &
\hspace{1cm} \rff \equiv \sd_{-}\;  \frac{8}{9}\; 
\tilde{\gamma}_0^2 \frac{\pi \beta_A^{3+2\delta}}{\Gamma(\frac{3}{2}+\delta)} 
\nonumber\\ & & 
\hspace{1cm} \breve{g}_{n} = \int_{0}^{\infty} dr\; 
r^{2+\delta}\;  j_{n}(\frac{1}{2} p_{B}r)\; 
\exp (- ( 2 \beta_{A}^{2} + \beta^{2}) \frac{r^{2}}{4})
\eeqna
where $\delta = 0.62$ \cite{page95light}, and $j_n, \; \Gamma$ are the spherical Bessel and  
gamma functions. Utilizing Eqs. \ref{fed}, \ref{hy} and \ref{wid} we obtain the
production strengths of hybrid mesons in Table \ref{table2} by first estimating the widths $\Gamma_{+}$ for constructive interference, and then correcting for the suppression caused by
destructive interference.

\begin{figure}[t]
\plabel{figty}
\caption{The ``lifting of suppression''. This is the ratio of $\protect\sd_{-}$
(Eq. \protect\ref{hybr}) to what would have been obtained
for $\protect\sd_{-}$ if the sign of $\protect\sff_{BC}$ in Eq. 
\protect\ref{fed} was positive. 
The ratio indicates the size of the actual destructive interference in
hybrid meson
two photon coupling, relative to the size of hybrid
meson two photon coupling that would result in the hypothetical case of constructive interference.
The lifting of suppression is plotted as a function of $Q \equiv \protect\sqrt{-q_1^2}$ 
($q_1^2 \leq 0$) where $q_2^2 = 0$. The case when one of the photon 
momenta is real
(in this case $q_2$) yields the maximal lifting of suppression as long as the 
physical production
constraint $q_i^2 \leq 0$ holds for the other photon. 
The graphs from highest to lowest lifting of suppression (at $Q=1$
GeV) are
for $I=0,\; I=1,\; s\bar{s}$ and $c\bar{c}$ flavours. The $c\bar{c}$ lifting of
suppression is maximal at $\approx 3.4$ GeV
and decreases at higher $Q$.}

\vspace{-1.5cm}

\centerline{\epsfysize=4in \epsffile{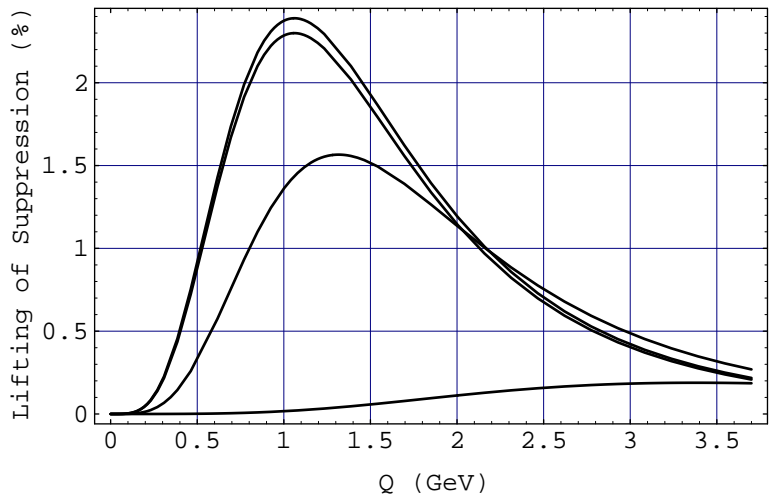}}

\vspace{-1.5cm}

\end{figure}

\begin{table}
\begin{center}
\caption{ Predicted upper bounds on the 
two photon ``widths'' $\Gamma_{max}$ (in eV)
of hybrids of various $J^{PC}$ and flavour.
These are obtained by multiplying the two photon widths 
$\Gamma_{+}$ (in eV) that would have been
attained if the sign in $\protect\sff_{BC}$ in Eq. \protect\ref{fed} 
was positive
instead of negative by the maximal ``lifting of suppression'' in Fig. 2, 
which 
is 2.3\%, 2.4\%, 1.6\%, 0.19\% for $I=1,\; I=0,\;  s\bar{s},\;  c\bar{c}$ 
respectively. 
$\Gamma_{max}$ is hence not a decay width, since the decay widths of
hybrid mesons are zero, but indicates the maximal production strength
of hybrid mesons in two photon processes.
For $u\bar{u}, \; s\bar{s}, \; c\bar{c}$ we use respectively 
$\beta_A=0.27, 0.30, 0.30$
GeV and $\beta = 0.37, 0.50, 0.57$ GeV. Hybrid masses are those of ref.
\protect\cite{page95light} i.e. $\approx$ 1.9, 2.1, 4.3 GeV for
$u\bar{u}, \; s\bar{s}, \; c\bar{c}$ hybrids.}
\plabel{table2}
\begin{tabular}{|l|c||l|l|l|l|}
\hline 
 \multicolumn{2}{|c||}{} & $I=1 $ & $I=0 $ & $s\bar{s} $ & $c\bar{c} $ \\
\hline \hline 
$\Gamma_{max}$ & $2^{-+}$ & 1 & 3  & 0.3 & 0.1\\
               & $1^{++}$ & 0.4 & 1 & 0.2 & 0.03\\
\hline 
$\Gamma_{+}$   & $2^{-+}$ & 40 & 140 & 20 & 50 \\
               & $1^{++}$ & 20 & 60 & 10 & 20 \\
\hline 
\end{tabular}
\end{center}
\end{table}

\section{Phenomenology and Conclusions \plabel{ph}}

We have exhibited a formalism to discuss two photon decays of
hybrid and conventional mesons through intermediate hadronic states.
The formalism respects Bose symmetry and ensures the consistency requirement that $J=1$ states
do not couple to two real photons. These are direct consequences of the 
creation of the $Q\bar{Q}$ pair with spin 1. The formalism also preserves the 
na\"{\i}ve expectation for the ratio of 
$I=1$ and $I=0$ amplitudes of $9/25$ for {\it both} mesons and hybrids demanded by the difference between
up and down quark electric charges, in accordance with
experiment\footnote{$\Gamma(a_2\rightarrow\gamma\gamma)/\Gamma(f_2\rightarrow\gamma\gamma)
\approx 9/25$ experimentally.}. 

Microscopic decay models \cite{ackleh96} have found
either dominant scalar confining interaction $^3P_0$, 
subdominant transverse one gluon exchange $^3S_1$ or highly suppressed
colour Coulomb one gluon exchange $^3P_0$, i.e. spin 1,  pair
creation for conventional mesons. If we assume that these decay
mechanisms also dominate for hybrids, 
we have demonstrated
that for both $^3P_0$ and $^3S_1$ pair creation Yang's theorem and Bose symmetry are satisfied,
and two photon decays of hybrid mesons vanish. 
These results are independent of detailed dynamics, but depends
on the valence quarks moving non--relativistically as the
only ``quenched'' quarks relevant to the connected decay, 
the absense of final state interactions, the spin assignment of a hybrid being that of
the adiabatic limit \cite{perantonis90,paton85}, and the photons coupling via
intermediate vector mesons. 
In the case of either the pair being created with spin 0 instead of
spin 1, or the hybrid spin assignments being the opposite to that of 
refs. \cite{perantonis90,paton85} hybrid widths may be non--vanishing.
 We caution that the breaking of 
assumptions made to derive the small hybrid production strengths
$\lapprox 0.03 - 3$ eV (Table \ref{table2}) can lead to more
significant production.
However, hybrid $\gamma\gamma$ widths are also small
in a relativistic model \cite{li92}, being 
$O(\alpha_S)$ suppressed relative to meson $\gamma\gamma$ widths.

Production strengths for hybrids are seen from Tables \ref{table1} and
\ref{table2} to be 
three orders of magnitude lower than for mesons,
and can hence be unpromising experimentally. 
This is especially relevant to the only low--lying exotic $J^{PC}$ accessable to
$\gamma\gamma$, i.e. $1^{-+}$, where there is additional suppression
in the formalism due to the photon coupling being restricted to longitudinal polarizations (see
Eq. \ref{hy}). Given the assumptions made, the main signature would
be vanishing production for $q^2_1 = q^2_2$.

Two photon collisions can be strong discriminators
in favour of the non--gluonic components of a physical state.
We have demonstrated\footnote{
Within this formalism, we find radially excited P and D--wave $Q\bar{Q}$
to possess $\sim 15\%$ the $\gamma\gamma$ widths of the ground states
in Table \protect\ref{table1} for realistic masses \protect\cite{isgur85}.
This is still substantial relative to the hybrid production in Table \protect\ref{table2}.
} this in a specific context for hybrid mesons, and
it is usually believed to be the case for glueballs \cite{berger87}.  For non--exotic $J^{PC}$ physical states are
expected \cite{page96rad}  to have both hybrid and conventional $Q\bar{Q}$ wave function components. 
Babar, Cleo II, LEP2 and LHC experiments 
should thus have considerable value in isolating 
substantial non--gluonic components in mixed states.
$f_0 (1500), \; f_J(1710)$ \cite{amsler96,li92}, 
$\rho(1450),\; \rho(1700)$, $\omega(1420),\; \omega(1600)$
\cite{page96rad,donnachie94}, $\psi(4040)$ and $\psi(4160)$ \cite{ono84,page95charm} have recently been suggested to be mixed.
In mixed states $\; \sin\phi\; |\mbox{non}-Q\bar{Q}\rangle +
\cos\phi\; |Q\bar{Q}\rangle \; $
and $\; \cos\phi\; |\mbox{non}-Q\bar{Q}\rangle - \sin\phi\; |Q\bar{Q}\rangle\; $
$\; \gamma\gamma$ collisions should isolate components
proportional to $\cos^2\phi$ and $\sin^2\phi$ respectively,
manifestly indicating the extent of
non--gluonic mixing by the strength of the two peaks appearing.

In the case of pure hybrid candidates with non--exotic $J^{PC}$,
$\gamma\gamma$ collisions offer the unique opportunity to isolate
and clearly distinguish the quarkonium partner, e.g. for the 1.8 GeV $0^{-+}$ isovector observed at
VES \cite{ves}.
Here the distinction is especially pronounced since $0^{-+}$ hybrid meson two photon production always vanishes in 
the flux--tube model (see Appendix \ref{ap1}).
VES possibly sees two
states, one a hybrid and the other a second radially excited $\pi$
\cite{page96rad,page96panic}. This should in principle show up as
two peaks in hadronic decay channels. Unfortunately, theoretical
predictions of hadronic decays 
of radially excited $Q\bar{Q}$ are highly
sensitive to parameter variations \cite{page96rad}, making comparisons to hybrid
decays difficult. Even though no radially
excited $Q\bar{Q}$ has been observed thus far in $\gamma\gamma$, 
it is within this formalism an exceptionally clean\footnote{
We expect a substantial $\gamma\gamma$ width of $2.3\pm 0.5$ keV
for a second radially excited $\pi$.} ``higher
quarkonium'' production mechanism, and a second radially excited $\pi$ may show up as one 
unambigous peak. Since the $\rho\omega$ decay of $0^{-+}$ has
been observed \cite{ves}, we expect the $\gamma\gamma$ collisions
production of the quarkonium component of $0^{-+}$ via intermediate 
$\rho,\omega$. 

Recently, an isoscalar $2^{-+}$ state at 1.875 GeV has been 
seen \cite{bugg96}, which appears to be the same as a state reported
previously by the Crystal Ball and Cello Collaborations
\cite{iso}. In addition, an isovector $2^{-+}$ at 1.8 GeV has
been reported by a number of collaborations \cite{page96rad}, most
recently VES \cite{ves}.
Detailed 
analysis \cite{page96rad} of these states leave it unsure whether they are radially
excited D--wave $Q\bar{Q}$ or hybrids. In addition, VES reported a 
2.2 GeV isovector $2^{-+}$ \cite{ves}, which has characteristics inconsistent
with expectations for a hybrid \cite{page95light}. We suggest that
$\gamma\gamma$ collisions
should be able to distinguish\footnote{
We expect radially excited $Q\bar{Q}\rightarrow\gamma\gamma$ widths of $0.02 {\lower.7ex\hbox{$\;\stackrel{\textstyle +0.01}{-0.02}\;$}}$ keV for the isovector, and $0.05{\lower.7ex\hbox{$\;\stackrel{\textstyle +0.06}{-0.03}\;$}}$ keV for the isoscalar.
} the non--gluonic content of these
states in the $1.8 - 2.2$ GeV region. 
The $\pi_2(1670)$ has been produced in $\gamma\gamma$ collisions
by the Crystal Ball and Cello Collaborations \cite{crystalcello}. 
However, in both of these experiments there 
are suggestive hints that there may be an isovector
contribution around 1.8 GeV, since the data appear to be
skewed towards the higher masses relative to simple Breit Wigner and
PDG values. Moreover, one expects
that the isoscalar $2^{-+}$ may also appear in $\gamma\gamma$ collisions, since
$\Gamma(\gamma\gamma\rightarrow I=0)$ $>$ $\Gamma(\gamma\gamma\rightarrow\ I=1)$.
Evidence for the isoscalar has been presented in ref. \cite{iso}.

We have seen that two photon collisions act as powerful discriminators between gluonic
and non--gluonic components and may considerably advance the isolation of gluonic
forms of matter, underlining the experimental potential of two photon physics.

\vspace{.5cm}

Helpful discussions with T. Barnes, A. Donnachie, J. Forshaw,
S.A. Sadovsky, P. Sutton, R. Williams and especially F.E. Close are acknowledged.

\appendix

\section{Appendix: Why Hybrid $0^{-+}$ Coupling to Two Vector Mesons Vanishes \protect\plabel{ap1}}

A general decay configuration can be rotated to a space fixed configuration
by using Appendix 1 of ref. \cite{paton85}

\beqn \psi(\br_H) = \sd^{L_H}_{M_{L}^{H} \Lambda_H} (\phi,\theta,\alpha)
\;\psi(\br_H^f) \eeqn
where state H  has orbital angular momentum $L_H$,
orbital angular momentum projection $M_{L}^{H}$, angular momentum $\Lambda_H$ around
the $Q\bar{Q}$ axis $\br_H$, and $\br_H^f$ indicates the axis in the fixed configuration.
H labels any of the states A, B or C.
Similarly, for the decay operator
\beqn \bsigma \cdot \bnabla = \sigma^{\sigma} \nabla_{\sigma}
= \sd^{1\ast}_{\sigma\mu}  (\phi,\theta,\alpha) \; \sigma^{\sigma} \nabla_{\mu}^f \eeqn 
where $\sigma$ denotes spherical basis coordinates.

Hence, focussing on the $\theta,\phi$ dependence of the flux--tube model decay amplitude
\cite{page95light}
$\propto \; \langle BC |\bsigma \cdot \bnabla | A\rangle$, we have
\beqn  \plabel{ang} \int\sin \theta \; d\theta \; d\phi \;  d\alpha \; \sd^{L_A}_{M_{L}^{A} \Lambda_A} (\phi,\theta,\alpha)\; 
\sd^{1\ast}_{\sigma\mu} (\phi,\theta,\alpha) \; 
\sd^{L_B\ast}_{M_{L}^{B} \Lambda_B} (\phi,\theta,\alpha)
\; \sd^{L_C\ast}_{M_{L}^{C} \Lambda_C} (\phi,\theta,\alpha)\;  
e^{\frac{i}{2} \bp_B \cdot \br} \eeqn
where the fixed configuration has been suppressed.
On $\phi$--integration the integral is only non--zero when
$M_L^A=\sigma+M_L^B+M_L^C$ which indicates conservation of
angular momentum projection. Similarly, on $\alpha$--integration,
$\mu = \Lambda_A - \Lambda_b - \Lambda_C$, which indicates that pair
creation absorbs the angular momentum in the fixed configuration.

For hybrid $0^{-+} \leftrightarrow VV$ we note that $L_B=L_C=0$, and
$M_L^A = \sigma$ by the first conservation rule.
Eq. \ref{ang} thus becomes
\beqn \plabel{pb} \sm_{M_{L}^A}\; (\bp_B) \sim \int  \sin \theta \; d\theta \; d\phi \; \sd^{L_A}_{M_{L}^{A} \Lambda_A} (\phi,\theta,-\phi)\; 
\sd^{1\ast}_{M_{L}^{A} \mu} (\phi,\theta,-\phi) \;  e^{\frac{i}{2} \bp_B\cdot \br} \eeqn
where we adopted the convention of ref. \cite{paton85}, setting
$\alpha = -\phi$ without loss of generality. Note that when 
$\bp_B \rightarrow -\bp_B$ Eq. \ref{pb} becomes
\beqn   \sm_{M_{L}^A} \; (-\bp_B) \sim \int\sin \theta \; d\theta \; d\phi \; \sd^{L_A}_{-M_{L}^{A} \Lambda_A} (\phi,\theta,-\phi)\; 
\sd^{1\ast}_{-M_{L}^{A} \mu} (\phi,\theta,-\phi) \;  e^{\frac{i}{2}  \bp_B \cdot\br} \eeqn
where we exchanged variables $\br \rightarrow -\br, \; \theta \rightarrow \pi - \theta, \phi \rightarrow \phi + \pi$ and used a property of the $\sd$--functions. It follows that $\sm_{0}$ and $\sm_1 + \sm_{-1}$ are invariant under
$\bp_B \rightarrow -\bp_B$. But from simple Clebsch--Gordon coefficients
for $0^{-+} \leftrightarrow VV$ we have that the 
$M_J^A=0,\;  M_J^B=\pm 1,\;  M_J^C=\mp 1$ amplitude is $-\frac{1}{\sqrt{2}} \sm_0$ and
the $M_J^A=0,\;  M_J^B=0,\;  M_J^C=0$ amplitude is $\frac{1}{\sqrt{2}} 
(\sm_1-\sm_0+\sm_{-1})$, both of which are accordingly invariant under
$\bp_B \rightarrow -\bp_B$. But since $0^{-+} \leftrightarrow VV$ is 
a P--wave decay the amplitude should be odd under $\bp_B \rightarrow -\bp_B$.
Hence the amplitude, and also the production and decay $0^{-+} \leftrightarrow VV$ vanishes.

To derive vanishing $0^{-+} \leftrightarrow VV$, we assumed
non--relativistically moving valence quarks,
with hybrid $0^{-+}$ having the spin assignment of refs. \cite{perantonis90,paton85} and decaying via $^3P_0$ OZI allowed pair creation. The argument does not
go through for $^3S_1$ pair creation.

\section{Appendix: Hybrid $\rightarrow$ 2S + 1S Mesons \protect\plabel{ap2}}

The decay amplitude for $hybrid \rightarrow 1S + 1S$ was displayed in
Eq. 21 of ref. \cite{page95light}. The amplitude for $hybrid \rightarrow 2S + 1S$ can
be obtained from it by noting that the 2S S.H.O. wave function equals
\beqn
\psi_{2S}(r) = 2\sqrt{\frac{2}{3}} \beta_{B}^{2}\; \frac{d}{d \beta_{B}^{2}}
\psi_{1S}(r)
\eeqn
The derivative operator can now be pulled in front of the overall
decay amplitude (Eq. 3 of ref. \cite{page95light}) and can hence be applied  to the
$hybrid \rightarrow 1S + 1S$ amplitude (Eq. 21 of ref. \cite{page95light}).
Noting that the amplitude for hybrid $\rightarrow$ 1S + 1S
is proportional to $\Delta = \beta_{B}^{2}-\beta_{C}^{2}$,
we first perform the differentiation and then set $\beta_{B}=\beta_{C}$
as a first orientation. Doing so sets $\Delta=0$, so the only contributing
term will be from the derivative acting on $\Delta$. This yields a    
$hybrid \rightarrow 2S + 1S$ amplitude of 

\beqn
2\sqrt{\frac{2}{3}} \beta_{B}^{2} \times (\; \mbox{Eq. 21 of ref. \cite{page95light}}\; ) / \Delta
\eeqn
when we restrict $\beta_{B}=\beta_{C}$. 

\section{Appendix: Width and Production Cross--Section \protect\plabel{ap3}}

Real photons can be distinguished according to the hemisphere in which they are
observed in an experimental apparatus, thus only appearing in a solid
angle $2\pi$, leading to the width relation

\beqna \plabel{wid} \lefteqn{\hspace{-1cm} \Gamma = \frac{1}{2J_A+1}\int (2\pi)^4\; \delta^4(q_1+q_2)\; \frac{d^3 q_1}{(2\pi)^3} 
\frac{d^3 q_2}{(2\pi)^3} \sum_{M_J^A,M_S^B,M_S^C} |\sm |^2 \nonumber } \\ & & 
= \frac{m_A^2}{16\pi (2J_A+1)} \sum_{M_J^A,M_S^B,M_S^C} |\sm |^2 \eeqna
where $m_A$ is the mass of state A.


\end{document}